\documentstyle[aps,multicol,epsf,epsfig]{revtex}
\newcommand{\ket}[1]{|{#1}\rangle}			
\newcommand{\bra}[1]{\langle{#1}|}			
\newcommand{\ketbra}[3]{\ket{#1}_{#3}\bra{#2}}

\newcommand{\sech}{\mathop{\rm sech}\nolimits} 

\begin{document}
\draft
\title{A Concentration/Purification Scheme for Two Partially Entangled 
Photon Pairs }	
\author{Takashi Yamamoto, Masato Koashi, and Nobuyuki Imoto}
\address{
CREST Research Team for Interacting Carrier Electronics, 
School of Advanced Sciences,\\ The Graduate University for 
Advanced Studies (SOKEN), Hayama, Kanagawa, 240-0193, Japan 
}
\maketitle

\begin{abstract}
An experimental scheme for concentrating entanglement in 
partially entangled photon pairs is proposed. 
In this scheme, two separated parties obtain one maximally
entangled photon pair from previously shared two partially entangled 
photon pairs by local operations and classical communication. 
A practical realization of the proposed scheme is discussed, which uses 
imperfect photon detectors and 
spontaneous parametric down-conversion as a photon source. 
This scheme also works for purifying a class of mixed states.
\pacs{PACS numbers: 03.67.-a 42.50.-p}
\end{abstract}

\begin{multicols}{2}

\section{INTRODUCTION}

In many applications in 
quantum information processing, such as quantum teleportation 
\cite{Bennett00,Zeilinger01,Kimble01} 
and entanglement based quantum key distribution 
\cite{Ekert01,QKD}, 
it is essential that two separated parties, Alice and Bob, share 
the maximally entangled particles in advance. 
Practically, a quantum channel, to be used to distribute 
the pairs, is usually noisy. 
It is thus important that Alice and Bob share 
maximally entangled pairs even through such channels. 
For that purpose, entanglement concentration \cite{Bennett01} and purification 
(or distillation) \cite{purification} have been originally proposed. 
In these schemes, previously shared less entangled pairs 
can be transformed into a smaller number of maximally entangled 
pairs by local operations and classical communication (LOCC). 
Until today, many schemes to obtain maximally entangled particles by LOCC 
have been proposed \cite{Knight01,purification02,purification03,Morikoshi}. 

In this paper we propose an experimentally feasible 
concentration/purification scheme, 
in which a maximally entangled photon pair is obtained from two photon 
pairs in identical partially entangled states. 
The basic idea of this paper is based on the concentration scheme 
proposed by Bennett, {\it et al.} \cite{Bennett01}. 
In their proposal, 
Alice or Bob performs a collective measurement 
for the joint state of $n$ pairs of particles (called 
as the Schmidt projection method), then they convert the projected state into 
a smaller number of maximally entangled pairs. 
For polarization entangled photons, however, 
the Schmidt projection method is difficult to perform because collective and 
non-destructive measurements for photons are not feasible today. 
In our scheme, Alice and Bob use only 
linear optical elements and photon detectors, in which destructive detection 
of two photons realizes the required projection and the conversion 
at the same time. 
In a similar scheme \cite{Knight01}, 
which uses entanglement swapping for two pairs of 
entangled photons, 
it is assumed that initially Alice has both photons of one pair and Alice 
and Bob share photons of the other pair. 
In our scheme, in contrast, we assume that the two pairs are 
distributed in the same way, namely, Alice obtains one member of 
each photon pair, and Bob obtains the other member of each photon pair, 
as shown in Fig.~\ref{fig:setup}. This feature makes the proposed scheme 
applicable to quantum channels with unknown fluctuations, namely 
the proposed scheme also works for purifying a class of mixed states. 
In the following, therefore, we use ^^ ^^ purification'' instead of 
^^ ^^ concentration/purification'' for simplicity. 

This paper is organized as follows. 
In Sec.\ \ref{sec:concept}, we explain our purification scheme 
in an ideal situation. 
In Sec.\ \ref{sec:imperfect detection}, we discuss two types of 
imperfect detectors and analyze the state after the purification. 
In Sec.\ \ref{sec:practical}, 
we consider the use of 
spontaneous parametric down-conversion (PDC) as a photon pair source and 
the effect of dark counts of the detectors. Finally, 
we discuss in Sec.\ \ref{sec:discussion}
the required property of fluctuating quantum channels for our scheme 
to be applicable. 

\section{Basic Idea }\label{sec:concept}

\begin{figure}[t]
 \centerline{\epsfig{width=8cm,file=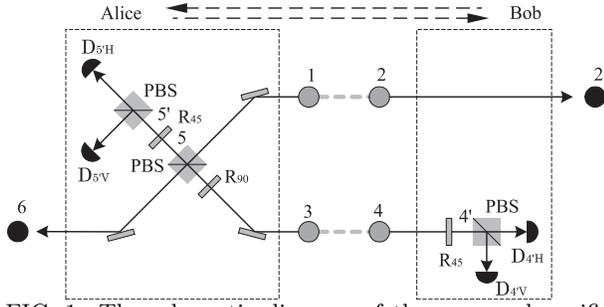}}
 \caption
 {The schematic diagram of the proposed purification scheme. 
 Polarization beam splitters (PBS) transmit ${\rm H}$ photons and 
 reflect ${\rm V}$ photons. $\lambda/2$ wave plates ${\rm R_{45}}$ and 
 ${\rm R_{90}}$ 
 rotate the polarization by $45^\circ$ and $90^\circ$ respectively.
 }
 \label{fig:setup}
\end{figure}

In this section, we show how the two separated parties, Alice and Bob, 
can purify a maximally entangled photon pair from two identical partially 
entangled photon pairs by LOCC. 
Let us assume that Alice and Bob are given two pairs of photons in the 
following polarization entangled states 
(we will describe a method creating this state in Sec.\ \ref{sec:practical})
\begin{eqnarray}
\ket{\alpha,\beta}_{12}\ket{\alpha,\beta}_{34}
&\equiv&
(\alpha\ket{1}_{1{\rm H}}\ket{1}_{2{\rm H}}+\beta\ket{1}_{1{\rm V}}\ket{1}_{2{\rm V}}) \nonumber \\
& &\otimes(\alpha\ket{1}_{3{\rm H}}\ket{1}_{4{\rm H}}+\beta\ket{1}_{3{\rm V}}\ket{1}_{4{\rm V}}), 
\label{eq:1}
\end{eqnarray}
where $\alpha$ and $\beta$ are complex numbers satisfying 
$|\alpha|^2+|\beta|^2=1$ and $\ket{n}$ is 
the normalized $n$-photon number state. 
The subscript numbers represent the spatial modes, and 
${\rm H}$ and ${\rm V}$ represent 
horizontal and vertical polarization modes, respectively. 
As shown in Fig.~\ref{fig:setup}, Alice receives photons in modes $1$ and $3$, 
and Bob receives photons in modes $2$ and $4$. 
For simplicity, we omit the modes in the vacuum, 
using abbreviations such as 
$\ket{1}_{1{\rm H}}\ket{1}_{2{\rm H}}\ket{0}_{1{\rm V}}\ket{0}_{2{\rm V}} 
\to \ket{1}_{1{\rm H}}\ket{1}_{2{\rm H}}$. 
Alice and Bob can transform these photons into a maximally 
entangled photon pair 
in modes 6 and 2, in the following way. 
\noindent
Eq.~(\ref{eq:1}) is expanded as 
\begin{eqnarray}
\alpha^2\ket{1}_{1{\rm H}}\ket{1}_{3{\rm H}}\ket{1}_{2{\rm H}}\ket{1}_{4{\rm H}}
+\beta^2\ket{1}_{1{\rm V}}\ket{1}_{3{\rm V}}\ket{1}_{2{\rm V}}\ket{1}_{4{\rm V}}  \nonumber \\
+\alpha\beta(\ket{1}_{1{\rm H}}\ket{1}_{3{\rm V}}\ket{1}_{2{\rm H}}\ket{1}_{4{\rm V}}
+\ket{1}_{1{\rm V}}\ket{1}_{3{\rm H}}\ket{1}_{2{\rm V}}\ket{1}_{4{\rm H}}).
\label{eq:2}
\end{eqnarray}
Note that the third and forth terms in Eq.~(\ref{eq:2}) have 
the same coefficients, 
$\alpha\beta$. Alice rotates 
the polarization of the photon in mode 3 by $90^\circ$ using 
a ${\rm \lambda}/2$ wave plate $({\rm R}_{{\rm 90}})$
and sends it to one port of a polarization beam splitter (PBS). 
The photon in mode 1 is sent to another port of the PBS.
After the PBS, 
the state of Eq.~(\ref{eq:2}) is transformed into 
\begin{eqnarray}
& &\alpha^2\ket{1}_{6{\rm H}}\ket{1}_{6{\rm V}}\ket{1}_{2{\rm H}}\ket{1}_{4{\rm H}}
+\beta^2\ket{1}_{5{\rm V}}\ket{1}_{5{\rm H}}\ket{1}_{2{\rm V}}\ket{1}_{4{\rm V}} \nonumber \\
& &+\alpha\beta(\ket{1}_{5{\rm H}}\ket{1}_{6{\rm H}}\ket{1}_{2{\rm H}}\ket{1}_{4{\rm V}}
+\ket{1}_{5{\rm V}}\ket{1}_{6{\rm V}}\ket{1}_{2{\rm V}}\ket{1}_{4{\rm H}}). 
\label{eq:3}
\end{eqnarray}
Note that there are two photons in the same spatial modes 
for the first two terms.
Alice and Bob rotate the polarization of their photons in modes 5 and 4 by 
$45^\circ$ using ${\rm \lambda}/2$ wave plates $({\rm R}_{{\rm 45}})$. 
These transformations are expressed by 
\begin{eqnarray}
\ket{1}_{k{\rm H}}\to 
\frac{1}{\sqrt{2}}(\ \ket{1}_{k^\prime {\rm H}}+\ket{1}_{k^\prime {\rm V}}\ ),
\label{eq:4}\\
\ket{1}_{k{\rm V}}\to 
\frac{1}{\sqrt{2}}(\ \ket{1}_{k^\prime {\rm H}}-\ket{1}_{k^\prime {\rm V}}\ ),
\label{eq:5}
\end{eqnarray}
and 
\begin{eqnarray}
\ket{1}_{k{\rm H}}\ket{1}_{k{\rm V}}\to 
\frac{1}{\sqrt{2}}(\ \ket{2}_{k^\prime {\rm H}}-\ket{2}_{k^\prime {\rm V}}\ ),\label{eq:5'}
\end{eqnarray}
where $k=4,5$. The state of 
Eq.~(\ref{eq:3}) is then transformed into 
\begin{eqnarray}
\ket{\Psi}
&=&\frac{\alpha^2}{\sqrt{2}}\ket{0}_{5^\prime}
  (\ket{1}_{4^\prime {\rm H}}+\ket{1}_{4^\prime {\rm V}} )
  \ket{1}_{6 {\rm H}}\ket{1}_{6 {\rm V}}\ket{1}_{2 {\rm H}} \nonumber \\
& &+\frac{\beta^2}{2}
  (\ket{2}_{5^\prime {\rm H}}\ket{1}_{4^\prime {\rm H}} 
       -\ket{2}_{5^\prime {\rm H}}\ket{1}_{4^\prime {\rm V}} 
	\nonumber \\
    & & 
	-\ket{2}_{5^\prime {\rm V}}\ket{1}_{4^\prime {\rm H}} 
       +\ket{2}_{5^\prime {\rm V}}\ket{1}_{4^\prime {\rm V}} 
  )\ket{0}_{6}\ket{1}_{2{\rm V}} \nonumber \\
& &+\frac{\alpha\beta}{\sqrt{2}}
  (\ket{1}_{5^\prime {\rm H}}\ket{1}_{4^\prime {\rm H}}\ket{\Phi^{(+)}}_{62} 
    -\ket{1}_{5^\prime {\rm H}}\ket{1}_{4^\prime {\rm V}}\ket{\Phi^{(-)}}_{62} 
	\nonumber \\
    & &
	+\ket{1}_{5^\prime {\rm V}}\ket{1}_{4^\prime {\rm H}}\ket{\Phi^{(-)}}_{62} 
    -\ket{1}_{5^\prime {\rm V}}\ket{1}_{4^\prime {\rm V}}\ket{\Phi^{(+)}}_{62} 
  ), \label{eq:5A}
\end{eqnarray}
where $\ket{\Phi^{(\pm )}}_{62}\equiv 1/\sqrt{2} 
(\ \ket{1}_{6{\rm H}}\ket{1}_{2{\rm H}}\pm \ket{1}_{6{\rm V}}\ket{1}_{2{\rm V}}\ )$ 
is the state of the maximally entangled photon pair. 
If Alice and Bob detect a single photon at ${\rm D}_{5^\prime {\rm H}}$ and 
${\rm D}_{4^\prime {\rm H}}$ (or ${\rm D}_{5^\prime {\rm V}}$ and ${\rm D}_{4^\prime {\rm V}}$) and 
the state is projected to 
$\ket{1}_{5^\prime {\rm H}}\ket{1}_{4^\prime {\rm H}}\ket{\Phi^{(+)}}_{62}$
(or $\ket{1}_{5^\prime {\rm V}}\ket{1}_{4^\prime {\rm V}}\ket{\Phi^{(+)}}_{62}$), 
they can share a maximally entangled photon pair in the state 
 $\ket{\Phi^{(+)}}_{62}$. 
If they detect a single photon at 
${\rm D}_{5^\prime {\rm H}}$ and ${\rm D}_{4^\prime {\rm V}}$ 
(or ${\rm D}_{5^\prime {\rm V}}$ and ${\rm D}_{4^\prime {\rm H}}$), they 
receive a maximally entangled photon pair in the state 
$\ket{\Phi^{(-)}}_{62}$. 
In this case, they can easily 
transform it into the form of $\ket{\Phi^{(+)}}_{62}$. 
Therefore the probability to share a maximally entangled photon pair 
in the state $\ket{\Phi^{(+)}}_{62}$ is $2|\alpha \beta|^2$. 

In this scheme, Alice and Bob need not know the values of 
$\alpha$ and $\beta$. 
Suppose that they receive the photons in a mixed state written as 
\begin{eqnarray}
\rho = \int P(\alpha,\beta)\ketbra{\alpha,\beta}{\alpha,\beta}{12}
	 \otimes \ketbra{\alpha,\beta}{\alpha,\beta}{34}
	d^2\alpha d^2\beta , \label{eq:mixture}
\end{eqnarray}
where $P(\alpha,\beta)$ is the probability distribution of 
their receiving the photon pairs in the state 
$\ket{\alpha,\beta}_{12}\ket{\alpha,\beta}_{34}$. 
In this case, the state of the photons just before 
the detection becomes a mixture of Eq.~(\ref{eq:5A}) with various values 
of $\alpha $ and $\beta $. 
They can, nevertheless, obtain a maximally entangled photon 
pair with the probability 
$\int 2|\alpha \beta|^2P(\alpha,\beta)d^2\alpha d^2\beta $ 
after they detect a single photon in modes $5^\prime $ and $4^\prime $. 
Since they can share a maximally entangled photon pair 
from pairs in a mixed state, 
this scheme can be called as entanglement purification. 

\section{purification using imperfect detection}
\label{sec:imperfect detection}

In this section, we study the property of output states in modes 6 and 2 
when detectors with a quantum efficiency $\eta$ are used. We consider 
two kinds of detectors; conventional photon detectors and 
single photon detectors. 
Conventional photon detectors (e.~g., EG\&G SPCM) 
cannot distinguish a single photon from two or more photons. 
Single photon detectors, which were recently demonstrated experimentally, 
can distinguish a single photon from two or more photons \cite{Takeuchi01}. 
In the following, we investigate the influence of the quantum efficiency 
on the output states in modes 6 and 2, and show that Alice and Bob receive 
a mixture of $\ket{\Phi^{(+)}}_{62}$ and $\ket{0}_{6}\ket{1}_{2{\rm V}}$ 
unless they use single photon detectors with a unit quantum efficiency. 

Consider a photon detector with a quantum efficiency $\eta$, 
which can distinguish any number of photocounts. 
Positive-operator-valued-measure (POVM) elements \cite{Peres}
of finding $n$ photocounts can be written as 
\cite{Barnett01}
\begin{eqnarray}
  \Pi_n=\sum\limits_{m=n}^\infty \eta^n(1-\eta)^{m-n}C^m_n\ket{m}_{}
  \bra{m}, 
\end{eqnarray}
where $C^m_n$ is the binomial coefficient, and 
$\sum_{n=0}^{\infty}\Pi_n =1$. 
Using this POVM, we can obtain the expression for the POVM elements for 
a conventional photon detector and a single photon detector. 
The POVM elements for a conventional photon detector 
can be written as \cite{Braunstein01}
 \begin{eqnarray}
  \Pi_{{\rm c}0}=\Pi_{0}
   =\sum\limits_{m=0}^\infty (1-\eta)^{m}\ket{m}_{}\bra{m}, 
 \end{eqnarray}
 and 
 \begin{eqnarray}
  \Pi_{{\rm c}1}=1-\Pi_{0}
   =\sum\limits_{m=1}^\infty [1-(1-\eta)^{m}]\ket{m}_{}\bra{m}.
  \label{eq:6}
 \end{eqnarray}
Here $\Pi_{{\rm c}0}$ is the POVM element for no photocounts, 
and $\Pi_{{\rm c}1}$ is that for photocounts. 
The POVM elements for a single photon detector can be written as 
 \begin{eqnarray}
  \Pi_{{\rm s}0}&=&\Pi_{0}
    =\sum\limits_{m=0}^\infty (1-\eta)^{m}\ket{m}_{}\bra{m}, \\
  \Pi_{{\rm s}1}&=&\Pi_{1}=\sum\limits_{m=1}^\infty 
		m\eta(1-\eta)^{m-1}\ket{m}_{}\bra{m},
 \end{eqnarray}
 and 
 \begin{eqnarray}
  \Pi_{{\rm s}2}&=&1-\Pi_{0}-\Pi_{1} \nonumber \\
      &=&\sum\limits_{m=2}^\infty 
	[1-(1-\eta+m\eta)(1-\eta)^{m-1}]\ket{m}_{}\bra{m}. \label{eq:7}
 \end{eqnarray}
Here $\Pi_{{\rm s}0}$ is the POVM element for no photocounts, 
$\Pi_{{\rm s}1}$ is that for single photocounts, and $\Pi_{{\rm s}2}$ is that 
for multiple photocounts. 
Using these POVM elements, we can calculate the output states 
after the detection at imperfect detectors ${\rm D}_{5^\prime {\rm H}}$, 
${\rm D}_{4^\prime {\rm H}}$, ${\rm D}_{5^\prime {\rm V}}$, and ${\rm D}_{4^\prime {\rm V}}$. 

Let us first consider the case where Alice and Bob use conventional photon 
detectors. 
Suppose that a coincidence detection is obtained at 
detectors ${\rm D}_{5^\prime {\rm H}}$ and ${\rm D}_{4^\prime {\rm H}}$. In this case 
photons are not detected at the detectors ${\rm D}_{5^\prime {\rm V}}$ or ${\rm D}_{4^\prime {\rm V}}$.
The output state in modes 6 and 2 after this detection is calculated as 
 \begin{eqnarray}
 & &\rho^{{\rm c}}_{{\rm out}}=\frac{{\rm Tr}_{5^\prime ,4^\prime }
   [\Pi^{5^\prime {\rm H}}_{{\rm c}1}\Pi^{4^\prime {\rm H}}_{{\rm c}1}\ketbra{\Psi}{\Psi}{} ]}
   {{\rm Tr}[\Pi^{5^\prime {\rm H}}_{{\rm c}1}\Pi^{4^\prime {\rm H}}_{{\rm c}1}\ketbra{\Psi}{\Psi}{}]} 
   \nonumber \\
   &=&\frac{
   |\alpha|^2\ket{\Phi^{(+)}}_{62}\bra{\Phi^{(+)}}
   +(1-\frac{\eta}{2})|\beta|^2
   \ket{0}_{6}\bra{0}\otimes \ket{1}_{2{\rm V}}\bra{1}}
   {1-\frac{\eta}{2}|\beta|^2}, \label{eq:8}
 \end{eqnarray}
where superscripts of POVM elements represent the modes. 
Note that Eq.~(\ref{eq:8}) is a classical mixture of 
the desired state $\ket{\Phi^{(+)}}_{62}$ and an error state 
$\ket{0}_{6}\ket{1}_{2{\rm V}}$. 
The probability of the coincidence 
detection $P$ can thus be regarded as the sum of two probabilities 
$P_{\rm s}$ and $P_{\rm e}$, where $P_{\rm s}$ is the probability of 
obtaining a photon pair in the state $\ket{\Phi^{(+)}}_{62}$ 
and $P_{\rm e}$ is the probability of 
obtaining a single photon in the state $\ket{0}_{6}\ket{1}_{2{\rm V}}$. 
These probabilities are calculated as 
$P={\rm Tr}[\Pi^{5^\prime {\rm H}}_{{\rm c}1}\Pi^{4^\prime {\rm H}}_{{\rm c}1}\ketbra{\Psi}{\Psi}{}]
=\eta^2 |\beta|^2 [ 2|\alpha|^2+(2-\eta)|\beta|^2 ]/4$, 
$P_{\rm s}=\eta^2 |\alpha\beta|^2 /2$ and 
$P_{\rm e}=\eta^2(2-\eta) |\beta|^4 /4$. 
The minimum value of $P_{\rm e}$ is $|\beta|^4 /4$. 
Alice and Bob can also obtain the output state $\rho^{{\rm c}}_{{\rm out}}$ 
when they obtain the other three combinations of coincidence, namely, 
(${\rm D}_{5^\prime {\rm V}}$, ${\rm D}_{4^\prime {\rm V}}$), 
(${\rm D}_{5^\prime {\rm H}}$, ${\rm D}_{4^\prime {\rm V}}$), 
and (${\rm D}_{5^\prime {\rm V}}$, ${\rm D}_{4^\prime {\rm H}}$). 
Therefore  the probability of obtaining the output state 
$\rho^{{\rm c}}_{{\rm out}}$ is $4P$. 

Similarly, in the case where Alice and Bob use single photon detectors, 
the output state in modes 6 and 2 after the detection is calculated as 
 \begin{eqnarray}
 & &\rho^{{\rm s}}_{{\rm out}}=\frac{Tr_{5^\prime ,4^\prime }
   [\Pi^{5^\prime {\rm H}}_{{\rm s}1}\Pi^{4^\prime {\rm H}}_{{\rm s}1}\ketbra{\Psi}{\Psi}{}]}
   {{\rm Tr}[\Pi^{5^\prime {\rm H}}_{{\rm s}1}\Pi^{4^\prime {\rm H}}_{{\rm s}1}\ketbra{\Psi}{\Psi}{}]} 
   \nonumber \\
   &=&\frac{
   |\alpha|^2\ket{\Phi^{(+)}}_{62}\bra{\Phi^{(+)}}
   +(1-\eta)|\beta|^2
   \ket{0}_{6}\bra{0}\otimes \ket{1}_{2{\rm V}}\bra{1}}
   {1-\eta|\beta|^2}. \label{eq:9}
 \end{eqnarray}
Note that Eq.~(\ref{eq:9}) is also a classical mixture of 
$\ket{\Phi^{(+)}}_{62}$ and $\ket{0}_{6}\ket{1}_{2{\rm V}}$. 
The probabilities $P$, $P_{\rm s}$, and $P_{\rm e}$ are calculated as 
$P={\rm Tr}[\Pi^{5^\prime {\rm H}}_{{\rm s}1}\Pi^{4^\prime {\rm H}}_{{\rm s}1}
\ketbra{\Psi}{\Psi}{}]=\eta^2 |\beta|^2 [ |\alpha|^2+(1-\eta)|\beta|^2 ]/2$, 
$P_{\rm s}=\eta^2 |\alpha\beta|^2 /2$, and 
$P_{\rm e}=\eta^2(1-\eta) |\beta|^4 /2$. 
Note that $P_{\rm s}$ is the same as in the case using 
the conventional photon detectors, but  $P_{\rm e}$ is different and 
its minimum value is $0$. 

The error in the output state $\rho^{{\rm c}}_{{\rm out}}$ or 
$\rho^{{\rm s}}_{{\rm out}}$ stems from the state 
$\ket{0}_{6}\ket{1}_{2{\rm V}}$ containing only one photon. 
Therefore, if Alice and Bob are allowed to perform the postselection, 
in which they select the events of the photocounts in modes 6 and 2, 
they can discard the events of error. 
In this situation, the types of detectors are not relevant, and the success 
probability is solely determined by the quantum efficiency $\eta$. 
\section{Implementation with a PDC source}
\label{sec:practical}
In this section, we consider 
the use of spontaneous parametric down-conversion as a photon source 
of the input states for the proposed purification scheme, 
and discuss the property of the output state. 
We also discuss the effect of the dark counts of the detectors. 
\subsection{Entangled photon pairs from PDC}
\begin{figure}[tb]
 \centerline {\epsfig{width=8cm,file=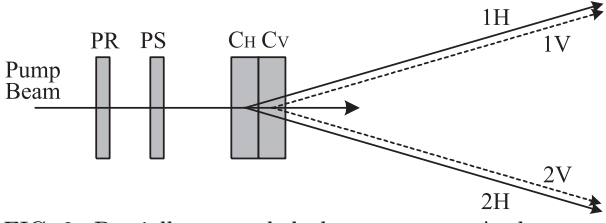}}
 \caption{Partially entangled photon source. 
 A photon pair in mode ${\rm H}$ and ${\rm V}$ is generated at 
 nonlinear crystals 
 ${\rm C_H}$ and ${\rm C_V}$ respectively. PR is a polarization rotator, and PS is 
 a phase shifter.}
 \label{fig:source}
\end{figure}

The partially entangled photon pair $\ket{\alpha,\beta}_{12}$ 
can be generated by 
pumping combined crystals, which is shown in 
Fig.~\ref{fig:source}\cite{Kwiat01}. 
The degree of entanglement can be continuously changed by rotating 
the polarization of the pump beam. 
The generated state $\ket{\Psi}_{12}$ can be written as 
$\ket{\Psi}_{12}=\ket{\Psi}_{12{\rm H}}\ket{\Psi}_{12{\rm V}}$, 
where $\ket{\Psi}_{12{\rm H}}$ and $\ket{\Psi}_{12{\rm V}}$ 
are the down-converted states generated from crystals ${\rm C_H}$ and ${\rm C_V}$, 
respectively, and are written as \cite{Yurke01Truax01}
 \begin{eqnarray}
 \ket{\Psi}_{12{\rm H}}=\sech |\gamma_{{\rm H}}| \displaystyle 
 \sum_{n=0}^{\infty}
 \biggl(\frac{\gamma_{\rm H}}{|\gamma_{{\rm H}}|}\tanh|\gamma_{{\rm H}}|
 \biggr)^n 
 \ket{n}_{1{\rm H}}\ket{n}_{2{\rm H}}  \label{eq:19}
 \end{eqnarray} 
 and 
 \begin{eqnarray}
 \ket{\Psi}_{12{\rm V}}=\sech |\gamma_{{\rm V}}| \displaystyle 
 \sum_{n=0}^{\infty}
 \biggl(\frac{\gamma_{\rm V}}{|\gamma_{{\rm V}}|}\tanh|\gamma_{{\rm V}}|
 \biggr)^n 
 \ket{n}_{1{\rm V}}\ket{n}_{2{\rm V}} \label{eq:20} 
 \end{eqnarray} 
\noindent
with $\gamma_{{\rm H}} \equiv |\gamma_{{\rm H}}|
e^{i(\phi_{{\rm p}}+\Delta \phi_{{{\rm p}}}/2)}$ and 
$\gamma_{{\rm V}} \equiv |\gamma_{{\rm V}}|e^{i(\phi_{{\rm p}}-\Delta 
\phi_{{{\rm p}}}/2)}$. 
Here, $\gamma_{{\rm H}} $ ($\gamma_{{\rm V}}$) is proportional to the  
complex amplitude of the classical pump beam for ${\rm C_H}$ (${\rm C_V}$). 
The phases of the pump beams for ${\rm C_H}$ and ${\rm C_V}$ 
are expressed by $\phi_{{\rm p}}+\Delta \phi_{{\rm p}}/2$ and 
$\phi_{{\rm p}}-\Delta \phi_{{\rm p}}/2$, respectively, where 
$\Delta \phi_{{\rm p}}$ is 
the phase difference between the two pump beams. 
The ratio of $|\gamma_{{\rm H}}|$ and $|\gamma_{{\rm V}}|$ 
can be controlled by 
rotating the polarization of the pump beam by the polarization rotator, PR, 
and $\Delta \phi_{{\rm p}}$ can be controlled by the phase shifter, PS. 
Using the following expressions, 
 \begin{eqnarray}
 \gamma \equiv \sqrt{\tanh^2|\gamma_{{\rm H}} |+\tanh^2|\gamma_{{\rm V}} |}, 
 \nonumber \\
 \alpha e^{i\phi_{{\rm p}}} \equiv 
   \frac{\gamma_{\rm H}}{|\gamma_{{\rm H}}|}\frac{\tanh|\gamma_{{\rm H}} |}
   {\gamma}, \nonumber \\
 \beta e^{i\phi_{{\rm p}}} \equiv
   \frac{\gamma_{\rm V}}{|\gamma_{{\rm V}}|}\frac{\tanh|\gamma_{{\rm V}} |}
   {\gamma}, \nonumber 
 \end{eqnarray}
 and 
 \begin{eqnarray}
 g \equiv \sech^2 |\gamma_{{\rm H}}|\sech^2 |\gamma_{{\rm V}}|
 =(1-\gamma^2|\alpha |^2)(1-\gamma^2|\beta |^2)
 ,\label{eq:21}
 \end{eqnarray}
we can write the state of the down-converted field as 
 \begin{eqnarray}
 \ket{\Psi}_{12}
   &=&\sqrt{g}
   \displaystyle \sum_{n=0}^{\infty} \displaystyle \sum_{m=0}^{\infty}
   (\gamma\alpha e^{i\phi_{{\rm p}}})^n
   (\gamma\beta e^{i\phi_{{\rm p}}})^m \nonumber  \\
   & &\times
   \ket{n}_{1{\rm H}}\ket{n}_{2{\rm H}}\ket{m}_{1{\rm V}}\ket{m}_{2{\rm V}}. \label{eq:22}
 \end{eqnarray}
Collecting the terms of the same total photon number, 
we can rewrite the state $\ket{\Psi}_{12}$ in the following form 
 \begin{eqnarray}
  \ket{\Psi}_{12}&=&\sqrt{g}
  ({}\ket{\Psi^{(0)}}_{12} \nonumber \\
  & &+\gamma e^{i\phi_{{\rm p}}}
  \ket{\Psi^{(1)}}_{12}+
         \gamma^2 e^{2i\phi_{{\rm p}}}\ket{\Psi^{(2)}}_{12}
  +\ldots ), \label{eq:23}
 \end{eqnarray}
where 
 \begin{eqnarray}
 \ket{\Psi^{(0)}}_{12}
 	&\equiv &\ket{0}_{1{\rm H}}\ket{0}_{2{\rm H}}\ket{0}_{1{\rm V}}\ket{0}_{2{\rm V}}, 
 	\label{eq:24_1}\\
 \ket{\Psi^{(1)}}_{12}&\equiv &
	\alpha
	\ket{1}_{1{\rm H}}\ket{1}_{2{\rm H}}\ket{0}_{1{\rm V}}\ket{0}_{2{\rm V}}\nonumber \\
	& &+\beta
	\ket{0}_{1{\rm H}}\ket{0}_{2{\rm H}}\ket{1}_{1{\rm V}}\ket{1}_{2{\rm V}}
	=\ket{\alpha ,\beta }_{12}, \label{eq:24_2}
 \end{eqnarray}
 and 
 \begin{eqnarray}
 \ket{\Psi^{(2)}}_{12}&\equiv &
	\alpha \beta
	\ket{1}_{1{\rm H}}\ket{1}_{2{\rm H}}\ket{1}_{1{\rm V}}\ket{1}_{2{\rm V}} \nonumber \\
  & &+\alpha^2
	\ket{2}_{1{\rm H}}\ket{2}_{2{\rm H}}\ket{0}_{1{\rm V}}\ket{0}_{2{\rm V}} \nonumber \\ 
  & &+\beta^2
	\ket{0}_{1{\rm H}}\ket{0}_{2{\rm H}}\ket{2}_{1{\rm V}}\ket{2}_{2{\rm V}}. \label{eq:24_3}
 \end{eqnarray}
\begin{figure}[tb]
 \centerline {\epsfig{width=8cm,file=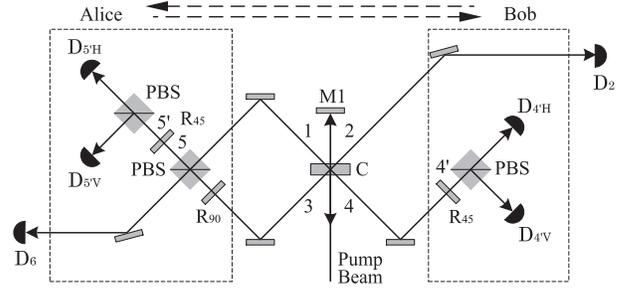}}
 \caption{Schematic of the purification procedure using 
 spontaneous parametric down-conversion as a photon source.
 \label{fig:PDCsetup}
 }
\end{figure}
\noindent
Note that $\ket{\Psi^{(0)}}_{12}$ and $\ket{\Psi^{(1)}}_{12}$ are 
normalized, but $\ket{\Psi^{(2)}}_{12}$ is not normalized. 

We will be able to obtain two photon pairs 
by pumping a nonlinear crystal twice by a short pulse as in 
Fig.~\ref{fig:PDCsetup}, 
as was used in several experiments \cite{Zeilinger01,Zeilinger02}.
The state $\ket{\Psi}_{1234}$ generated from this source can be expressed as 
\begin{eqnarray}
\ket{\Psi}_{1234}&=&\ket{\Psi}_{12}\ket{\Psi}_{34} \nonumber \\
  &=&g\Bigl[{}\ket{\Psi^{(0)}}_{12}\ket{\Psi^{(0)}}_{34} \nonumber \\
  & &+\gamma e^{i\phi_{{\rm p}}}( \ket{\Psi^{(1)}}_{12}\ket{\Psi^{(0)}}_{34} 
        +\ket{\Psi^{(0)}}_{12}\ket{\Psi^{(1)}}_{34} ) \nonumber \\ 
  & &+\gamma^2 e^{2i\phi_{{\rm p}}} 
  ( \ket{\Psi^{(1)}}_{12}\ket{\Psi^{(1)}}_{34} \nonumber \\ 
  & &+\ket{\Psi^{(2)}}_{12}\ket{\Psi^{(0)}}_{34} 
  +\ket{\Psi^{(0)}}_{12}\ket{\Psi^{(2)}}_{34} ) +\ldots \Bigr] 
  \nonumber \\
  &=&g({}\ket{\Psi^{(0)}}_{1234}+
       \gamma e^{i\phi_{{\rm p}}}\ket{\Psi^{(1)}}_{1234} \nonumber \\ 
  & &+\gamma^2 e^{2i\phi_{{\rm p}}} \ket{\Psi^{(2)}}_{1234} +\ldots ),
  		 \label{eq:25}
\end{eqnarray}
where $\ket{\Psi^{(0)}}_{1234}\equiv 
\ket{\Psi^{(0)}}_{12}\ket{\Psi^{(0)}}_{34}$, 
$\ket{\Psi^{(1)}}_{1234} \equiv \ket{\Psi^{(1)}}_{12}\ket{\Psi^{(0)}}_{34}
+\ket{\Psi^{(0)}}_{12}\ket{\Psi^{(1)}}_{34}$ and 
$\ket{\Psi^{(2)}}_{1234} \equiv \ket{\Psi^{(1)}}_{12}\ket{\Psi^{(1)}}_{34}
+\ket{\Psi^{(2)}}_{12}\ket{\Psi^{(0)}}_{34}+
\ket{\Psi^{(0)}}_{12}\ket{\Psi^{(2)}}_{34}$. 
In our scheme, Alice and Bob do not know the phase $\phi_{{\rm p}}$, so that 
the state received by them is the mixed state $\rho^{{\rm PDC}}_{1234}$ 
that is obtained by averaging Eq.~(\ref{eq:25}) over $\phi_{{\rm p}}$ as 
\begin{eqnarray}
\rho^{{\rm PDC}}_{1234}&=&g^2(\ketbra{\Psi^{(0)}}{\Psi^{(0)}}{1234}
	+\gamma^2\ketbra{\Psi^{(1)}}{\Psi^{(1)}}{1234} \nonumber \\ 
	& &+\gamma^4\ketbra{\Psi^{(2)}}{\Psi^{(2)}}{1234} +\ldots ).
\end{eqnarray}
In the following, we assume that $\gamma$ is small, so that 
we restrict the analysis up to $O(\gamma^4)$. 
\subsection{Purification using imperfect detection}
As shown in Fig.~\ref{fig:PDCsetup}, the state $\rho^{{\rm PDC}}_{1234}$ 
is transformed by the same operations described in Sec.\ \ref{sec:concept}. 
The term $\ket{\Psi^{(1)}}_{12}\ket{\Psi^{(1)}}_{34}$ becomes 
Eq.~(\ref{eq:5A}) and the other terms are calculated as 
\begin{eqnarray}
& &\ket{\Psi^{(1)}}_{12}\ket{\Psi^{(0)}}_{34} \nonumber \\
  & &\to \alpha\ket{1}_{6{\rm H}}\ket{1}_{2{\rm H}} \nonumber \\
	& &+\frac{\beta}{\sqrt{2}}
	( \ket{1}_{5^\prime {\rm H}}-\ket{1}_{5^\prime {\rm V}} )\ket{1}_{2{\rm V}}, 
\end{eqnarray}

\begin{eqnarray}
& &\ket{\Psi^{(0)}}_{12}\ket{\Psi^{(1)}}_{34} \nonumber \\
  & &\to \frac{\alpha}{\sqrt{2}}
	\ket{1}_{6{\rm V}}( \ket{1}_{4^\prime {\rm H}}+\ket{1}_{4^\prime {\rm V}} ) 
	\nonumber \\
  & &+\frac{\beta}{2}
	( \ket{1}_{5^\prime {\rm H}}+\ket{1}_{5^\prime {\rm V}})
	( \ket{1}_{4^\prime {\rm H}}-\ket{1}_{4^\prime {\rm V}}), 
\end{eqnarray}
\begin{eqnarray}
& &\ket{\Psi^{(2)}}_{12}\ket{\Psi^{(0)}}_{34}\nonumber \\
  & &\to \alpha^2
	\ket{2}_{6{\rm H}}\ket{2}_{2{\rm H}}  \nonumber \\
  & &+\frac{\beta^2}{2}
	(\ket{2}_{5^\prime {\rm H}}
	-\sqrt{2}\ket{1}_{5^\prime {\rm H}}\ket{1}_{5^\prime {\rm V}} 
	+\ket{2}_{5^\prime {\rm V}} ) \ket{2}_{2{\rm V}} \nonumber \\
  & &+\frac{\alpha \beta}{\sqrt{2}}
	(\ket{1}_{5^\prime {\rm H}}-\ket{1}_{5^\prime {\rm V}} ) 
	\ket{1}_{6{\rm H}}\ket{1}_{2{\rm H}}\ket{1}_{2{\rm V}}, 
\end{eqnarray}
and 
\begin{eqnarray}
& &\ket{\Psi^{(0)}}_{12}\ket{\Psi^{(2)}}_{34} \nonumber \\
  & &\to \frac{\alpha^2}{2}
	( \ket{2}_{4^\prime {\rm H}}
	+\sqrt{2}\ket{1}_{4^\prime {\rm H}}\ket{1}_{4^\prime {\rm V}}
	+\ket{2}_{4^\prime {\rm V}})\ket{2}_{6{\rm V}}  \nonumber \\
  & &+\frac{\beta^2}{4}
	( \ket{2}_{5^\prime {\rm H}}
	+\sqrt{2}\ket{1}_{5^\prime {\rm H}}\ket{1}_{5^\prime {\rm V}}
	+\ket{2}_{5^\prime {\rm V}}) \nonumber \\
  & &\qquad \otimes ( \ket{2}_{4^\prime {\rm H}}
	-\sqrt{2}\ket{1}_{4^\prime {\rm H}}\ket{1}_{4^\prime {\rm V}}
	+\ket{2}_{4^\prime {\rm V}})
	  \nonumber \\
  & &+\frac{\alpha \beta}{2}
	( \ket{1}_{5^\prime {\rm H}}+\ket{1}_{5^\prime {\rm V}})
	(\ket{2}_{4^\prime {\rm H}}-\ket{2}_{4^\prime {\rm V}})\ket{1}_{6{\rm V}}.  
    \label{eq:Psi34(2)}
\end{eqnarray}
Using these expressions, we can obtain the state 
$\rho^{{\rm PDC}}_{2 4^\prime 5^\prime 6}$ after transforming 
$\rho^{{\rm PDC}}_{1234}$. 

We calculate the output state in modes 6 and 2 by using a similar way 
as in Sec.\ \ref{sec:imperfect detection}. 
Let us consider the case where Alice and Bob use conventional photon 
detectors. Suppose that a coincidence detection is obtained at 
detectors ${\rm D}_{5^\prime {\rm H}}$ and ${\rm D}_{4^\prime {\rm H}}$. 
In contrast to the case in Sec.\ \ref{sec:imperfect detection}, 
the modes $5^\prime {\rm V}$ and $4^\prime {\rm V}$ are not always 
in the vacuum. 
If photocounts are recorded at detectors 
${\rm D}_{5^\prime {\rm V}}$ or ${\rm D}_{4^\prime {\rm V}}$, 
they obtain the vacuum in modes 6 and 2. 
It is thus better to discard such events in order to reduce errors. 
When the detectors 
${\rm D}_{5^\prime {\rm V}}$ and ${\rm D}_{4^\prime {\rm V}}$ 
record no photocounts, the output state in modes 6 and 2 after the detection 
is calculated as 
\begin{eqnarray}
 \rho^{{\rm c}}_{{\rm out}}&=&\frac{{\rm Tr}_{5^\prime ,4^\prime }
   [\Pi^{5^\prime {\rm H}}_{{\rm c}1}\Pi^{4^\prime {\rm H}}_{{\rm c}1}
	\Pi^{5^\prime {\rm V}}_{{\rm c}0}\Pi^{4^\prime {\rm V}}_{{\rm c}0}
    \rho^{{\rm PDC}}_{2 4^\prime 5^\prime 6}]}
   {{\rm Tr}[\Pi^{5^\prime {\rm H}}_{{\rm c}1}\Pi^{4^\prime {\rm H}}_{{\rm c}1}
             \Pi^{5^\prime {\rm V}}_{{\rm c}0}\Pi^{4^\prime {\rm V}}_{{\rm c}0}
             \rho^{{\rm PDC}}_{2 4^\prime 5^\prime 6}]} 
	\nonumber \\
   &=&\frac{1}{C^{{\rm c}}} 
	\Bigl\{ \ 8\gamma^2|\alpha|^2\ket{\Phi^{(+)}}_{62}\bra{\Phi^{(+)}}
	\nonumber \\ 
   & &\qquad 
	 +[4+(4-3\eta)^2\gamma^2|\beta|^2 ]
              \ket{0}_{6}\bra{0}\otimes\ket{0}_{2}\bra{0} \nonumber \\
   & &\qquad 
	+4(2-\eta)\gamma^2|\beta|^2\ket{0}_{6}\bra{0}\otimes\ket{1}_{2{\rm V}}\bra{1}
   	\nonumber \\
   & &\qquad 
	+4(2-\eta)\gamma^2|\alpha|^2
	\ket{1}_{6{\rm V}}\bra{1}\otimes\ket{0}_{2}\bra{0}
	\ \Bigr\}, \label{eq:32}
 \end{eqnarray}
where $C^{{\rm c}} = 4+4(4-\eta)\gamma^2|\alpha|^2
+(24-28\eta+9\eta^2)\gamma^2|\beta|^2$. 
Note that Eq.~(\ref{eq:32}) is also a classical mixture of 
$\ket{\Phi^{(+)}}_{62}$ and 
error states containing smaller number of photons. 
As in Sec.\ \ref{sec:imperfect detection}, we use the probabilities 
$P$, $P_{\rm s}$, and $P_{\rm e}$, but here we further decompose $P_{\rm e}$ 
as $P_{\rm e}=P^{(0)}_{\rm e}+P^{(1)}_{\rm e}$, where 
$P^{(0)}_{\rm e}$ is the probability of having 
the vacuum in modes 6 and 2, and $P^{(1)}_{\rm e}$ is that of 
having a photon either in mode 6 or 2. 
Each probability is expressed as 
 \begin{eqnarray}
   P&=&\eta^2g^2\gamma^2|\beta|^2C^{{\rm c}}/16, \nonumber \\
   P_{\rm s}&=&\eta^2g^2\gamma^4|\alpha\beta|^2/2, \nonumber\\
   P^{(0)}_{\rm e}&=&\eta^2g^2\gamma^2|\beta|^2
   		[ 4+(4-3\eta)^2\gamma^2|\beta|^2 ]/16, \nonumber 
 \end{eqnarray}
 and 
 \begin{eqnarray}
  P^{(1)}_{\rm e}&=&\eta^2(2-\eta)g^2\gamma^4|\beta|^2/4.
 \end{eqnarray}
In this case the minimum value of $P^{(0)}_{\rm e}$ and $P^{(1)}_{\rm e}$ are 
$g^2\gamma^2|\beta|^2 [ 4+\gamma^2|\beta|^2 ]/16$ 
and $g^2\gamma^4|\beta|^2/4$, respectively. 
If Alice and Bob do not discard the events when photocounts are recorded at 
detectors ${\rm D}_{5^\prime {\rm V}}$ or ${\rm D}_{4^\prime {\rm V}}$, 
$P^{(0)}_{\rm e}$ increases to $\eta^2g^2\gamma^2|\beta|^2
[ 4+(4-\eta)^2\gamma^2|\beta|^2 ]/16$ and the minimum value of 
$P^{(0)}_{\rm e}$ 
increases to $g^2\gamma^2|\beta|^2 [ 4+9\gamma^2|\beta|^2]/16$. 

Similarly, in the case where Alice and Bob use single photon detectors, 
the output state in modes 6 and 2 after the detection 
is calculated as 
 \begin{eqnarray}
 \rho^{{\rm s}}_{{\rm out}}&=&\frac{{\rm Tr}_{5^\prime ,4^\prime }
   [\Pi^{5^\prime {\rm H}}_{{\rm s}1}\Pi^{4^\prime {\rm H}}_{{\rm s}1}
     \Pi^{5^\prime {\rm V}}_{{\rm s}0}\Pi^{4^\prime {\rm V}}_{{\rm s}0}
    \rho^{{\rm PDC}}_{2 4^\prime 5^\prime 6}]}
   {{\rm Tr}[\Pi^{5^\prime {\rm H}}_{{\rm s}1}\Pi^{4^\prime {\rm H}}_{{\rm s}1}
             \Pi^{5^\prime {\rm V}}_{{\rm s}0}\Pi^{4^\prime {\rm V}}_{{\rm s}0}
             \rho^{{\rm PDC}}_{2 4^\prime 5^\prime 6}]} 
		\nonumber \\
   &=&\frac{1}{C^{{\rm s}}} 
      \Bigl\{\ 2\gamma^2|\alpha|^2\ket{\Phi^{(+)}}_{62}\bra{\Phi^{(+)}}
	\nonumber \\
   & &\qquad 
	+[ 1+4(1-\eta)^2\gamma^2 |\beta|^2 ]\ket{0}_{6}\bra{0}
	\otimes\ket{0}_{2}\bra{0} 
	\nonumber \\
   & &\qquad 
	+2(1-\eta)\gamma^2|\beta|^2\ket{0}_{6}\bra{0}\otimes\ket{1}_{2{\rm V}}\bra{1}
   	\nonumber \\
   & &\qquad 
	+2(1-\eta)\gamma^2|\alpha|^2\ket{1}_{6{\rm V}}\bra{1}
	\otimes\ket{0}_{2}\bra{0}
	\ \Bigr\}, \label{eq:33}
 \end{eqnarray}
where $C^{{\rm s}} = 1+2(2-\eta)\gamma^2|\alpha|^2
	+2(3-2\eta)(1-\eta)\gamma^2|\beta|^2$. 
Each probability is expressed as
 \begin{eqnarray}
   P&=&\eta^2g^2\gamma^2|\beta|^2C^{{\rm s}}/4, \nonumber \\
   P_{\rm s}&=&\eta^2g^2\gamma^4|\alpha\beta|^2/2, \nonumber\\
   P^{(0)}_{\rm e}&=&\eta^2g^2\gamma^2|\beta|^2
   		[ 1+4(1-\eta)^2\gamma^2|\beta|^2 ]/4, \nonumber 
 \end{eqnarray}
 and 
 \begin{eqnarray}
  P^{(1)}_{\rm e}&=&\eta^2(1-\eta)g^2\gamma^4|\beta|^2/2.
 \end{eqnarray}
In this case the minimum values of $P^{(0)}_{\rm e}$ and $P^{(1)}_{\rm e}$ 
are $g^2\gamma^2|\beta|^2 /4$ and $0$, respectively. 
Note that in comparison with the case using the conventional photon 
detectors, $P_{\rm s}$ is the same, but $P^{(0)}_{\rm e}$ and 
$P^{(1)}_{\rm e}$ are different.  
If Alice and Bob do not discard the events when photocounts are recorded at 
detectors ${\rm D}_{5^\prime {\rm V}}$ or ${\rm D}_{4^\prime {\rm V}}$, 
$P^{(0)}_{\rm e}$ increases to $\eta^2g^2\gamma^2|\beta|^2
[ 1+(2-\eta)^2\gamma^2|\beta|^2]/4$ and the minimum value of $P^{(0)}_{\rm e}$ 
increases to $g^2\gamma^2|\beta|^2[ 1+\gamma^2|\beta|^2]/4$. 

The error in the output state $\rho^{{\rm c}}_{{\rm out}}$ or 
$\rho^{{\rm s}}_{{\rm out}}$ stems from the states 
with smaller number of photons. 
Therefore, if Alice and Bob are allowed to perform the postselection, 
they can discard the events of error 
similarly to the case of the ideal photon pair source. 
In this situation, again, 
the types of detectors are not relevant and the success 
probability is solely determined by the quantum efficiency $\eta$. 
Moreover, they need not refer to the detectors 
${\rm D}_{5^\prime {\rm V}}$ and ${\rm D}_{4^\prime {\rm V}}$ 
because the vacuum is removed by the postselection. 
\subsection{The effect of dark counts}
When photon detectors have dark counts, the probability of 
error $P_{\rm e}$ 
increases, and the error cannot always be discarded even by 
the postselection.  
In the following, we derive the conditions that we can neglect 
the effect of dark counts. 

We assume that the dark counts are random detection events, namely, 
each event is uncorrelated to other dark or real counts. 
Let the mean number of dark counts during each run of the purification scheme 
be $\nu$ for each detector. We assume $\nu\ll 1$. 
Consider the case Alice and Bob 
obtain a fourfold coincidence detection at detectors 
${\rm D}_{5^\prime {\rm H}}$, ${\rm D}_{4^\prime {\rm H}}$, ${\rm D}_{6}$, 
and ${\rm D}_{2}$. 
The probability that all the four counts are caused by real photons is 
$P_0=O(\gamma^4)$. 
$\gamma^2$ is the generation probability of a photon pair. 
The probabilities $P_i$ that the fourfold 
coincidence detection includes $i$ dark counts are of the order 
 $P_1=O(\gamma^4\nu)$, $P_2=O(\gamma^2\nu^2)$, 
$P_3=O(\gamma^2\nu^3)$ and $P_4=O(\nu^4)$. 
To satisfy $P_0 \gg  P_i \quad (i=1,2,3,4)$, $\nu$ must satisfy 
$\nu^2/\gamma^2 \ll 1$. Therefore, the condition 
for the effect of dark counts to be negligible is $\nu\ll 1$ and 
$\nu^2/\gamma^2 \ll 1$. 

In a teleportation experiment \cite{Zeilinger01,Zeilinger03}, 
where a nonlinear crystal is pumped twice by a short pulse, 
$\gamma^2$ is of order $\sim \!\! 10^{-4}$. 
The conventional photon detectors (e.~g., EG\&G SPCM) typically have 
the dark count rates of order $100\ {\rm s}^{-1}$, which gives the value of 
$\nu \sim \!\! 10^{-6}$ for the coincidence time $\sim \!\! 10\ {\rm ns}$.
The single photon detectors \cite{Takeuchi01} have the dark count 
rates of order $10^4\ {\rm s}^{-1}$, which gives the value of 
$\nu \sim \!\! 10^{-4}$ 
for the coincidence time $\sim \!\! 10\ {\rm ns}$.
In both cases the effect of dark counts is negligible.

\section{Discussion and conclusion}\label{sec:discussion}
In the following we consider the required property of quantum channels for 
the proposed purification scheme to be applicable. 
Assume that two photon pairs 
are initially prepared in the state 
$\ket{\Phi^{(+)}}_{12}\ket{\Phi^{(+)}}_{34}$ 
and sent to Alice and Bob through noisy quantum channels. 
The quantum channels are assumed to have 
polarization-dependent transmissivities and are modeled by the state 
transformation 
\begin{eqnarray}
\ket{1}_{kL}\to 
(\ \mu_{kL}\ket{1}_{kL}+\sqrt{1-|\mu_{kL}|^2}\ket{1}_{\bar{k} L}\ ),
\end{eqnarray}
where $k=1,2,3,4$, $L={\rm H},{\rm V}$, and $\mu_{kL}$ is complex 
transmission coefficient. 
We introduce modes $\bar{k} L$ to model lossy channels. The coefficients 
$\mu_{kL}$ are fluctuating and we denote the average over the fluctuations 
as $\langle \cdots \rangle_{\mu}$. 
The state of photon pairs received by Alice and Bob is 
\begin{eqnarray}
\langle (1-P)\rho^{n\le 3}_{1234}+ 
	P\rho_{1234} \rangle_{\mu} , 
\end{eqnarray}
where $\rho^{n\le 3}_{1234}$ is the state containing 
less than four photons in total and 
\begin{eqnarray}
	P&\equiv &\frac{1}{4}(|\mu_{1{\rm H}}\mu_{2{\rm H}}|^2
			+|\mu_{1{\rm V}}\mu_{2{\rm V}}|^2) \nonumber \\
		& &\quad \times (|\mu_{3{\rm H}}\mu_{4{\rm H}}|^2
			+|\mu_{3{\rm V}}\mu_{4{\rm V}}|^2), \nonumber \\
	\rho_{1234}&\equiv &
	\ketbra{\alpha_{12},\beta_{12}}
	{\alpha_{12},\beta_{12}}{12} \nonumber \\
	 & &\quad \otimes \ketbra{\alpha_{34},\beta_{34}}
	 {\alpha_{34},\beta_{34}}{34}, \nonumber \\
	\alpha_{12}&\equiv &\frac{\mu_{1{\rm H}}\mu_{2{\rm H}}}
			{\sqrt{|\mu_{1{\rm H}}\mu_{2{\rm H}}|^2
				+|\mu_{1{\rm V}}\mu_{2{\rm V}}|^2}}, 
	\nonumber \\
	\beta_{12}&\equiv &\frac{\mu_{1{\rm V}}\mu_{2{\rm V}}}
			{\sqrt{|\mu_{1{\rm H}}\mu_{2{\rm H}}|^2
				+|\mu_{1{\rm V}}\mu_{2{\rm V}}|^2}}, 
	\nonumber \\
	\alpha_{34}&\equiv &\frac{\mu_{3{\rm H}}\mu_{4{\rm H}}}
			{\sqrt{|\mu_{3{\rm H}}\mu_{4{\rm H}}|^2
				+|\mu_{2{\rm V}}\mu_{4{\rm V}}|^2}}, \nonumber
\end{eqnarray}
and 
\begin{eqnarray}
	\beta_{34}&\equiv &\frac{\mu_{3{\rm V}}\mu_{4{\rm V}}}
			{\sqrt{|\mu_{3{\rm H}}\mu_{4{\rm H}}|^2
				+|\mu_{2{\rm V}}\mu_{4{\rm V}}|^2}}.
\end{eqnarray}

If the postselection is allowed, Alice and Bob can 
select two photon pairs in the state 
$\langle P \rho_{1234} \rangle_{\mu} /\langle P \rangle_{\mu}$. 
To purify the mixed state 
$\langle P \rho_{1234} \rangle_{\mu} /\langle P \rangle_{\mu}$, 
this mixed state must be written in the form of Eq.~(\ref{eq:mixture}). 
By comparing the matrix elements of these expressions, we obtain the condition 
for the purification as 
$\langle P |\alpha_{12}\beta_{34}-\beta_{12}\alpha_{34}|^2 
\rangle_{\mu}=0$. 
Using the following complex variable 
\begin{eqnarray}
	F &\equiv & \frac{\alpha_{12}\beta_{34}}{\beta_{12}\alpha_{34}} 
	  =	\frac{\mu_{1{\rm H}}\mu_{2{\rm H}}\mu_{3{\rm V}}\mu_{4{\rm V}}}
		{\mu_{1{\rm V}}\mu_{2{\rm V}}\mu_{3{\rm H}}\mu_{4{\rm H}}}, 
	\label{eq:F}
\end{eqnarray}
the condition for the purification 
becomes $F=1$. Even if $F\neq 1$, Alice and Bob can 
transform $F$ into 1 by introducing an additional attenuation and 
a phase shift as long as the value $F$ is constant. 
The fluctuations in the transmissivities of the quantum channels 
may be assumed to be independent for Alice's side and Bob's side. 
In this case we can introduce complex variables 
\begin{eqnarray}
	F_{\rm A} 
	\equiv \frac{\mu_{1{\rm H}}\mu_{3{\rm V}}}{\mu_{1{\rm V}}\mu_{3{\rm H}}}, 
	\label{eq:FA}
\end{eqnarray}
and 
\begin{eqnarray}
	F_{\rm B} 
	\equiv \frac{\mu_{2{\rm H}}\mu_{4{\rm V}}}{\mu_{2{\rm V}}\mu_{4{\rm H}}}, 
	\label{eq:FB}
\end{eqnarray}
where $F=F_{\rm A} F_{\rm B}$. 
Since $F_{\rm A}$ and $F_{\rm B}$ are independent, the condition for the purification is 
that $F_{\rm A}$ and $F_{\rm B}$ are constant. 

In the special cases where each pair is received as  a known pure state 
$\ket{\alpha_{12},\beta_{12}}\otimes \ket{\alpha_{34},\beta_{34}}$, 
the Procrustean method \cite{Bennett01} 
can be applied to each pair. 
In this method, Alice and Bob perform an additional 
polarization-dependent transformation to discard the extra probability of 
the larger term in the state $\ket{\alpha_{12},\beta_{12}}_{12}$. 
Since they manipulate one photon pair, 
this method is simpler than the proposed scheme to 
share maximally entangled state. 
To simplify our explanation, we consider the situations where Bob prepares 
the photon pairs 
and sends one member of each photon pair to Alice 
through quantum channels $1$ and $3$, namely 
$\mu_{2{\rm H}}=\mu_{2{\rm V}}=\mu_{4{\rm H}}=\mu_{4{\rm V}}=1$.
The Procrustean method is then applicable when the fluctuations 
$\mu_{1{\rm H}}$, $\mu_{1{\rm V}}$, $\mu_{3{\rm H}}$, and $\mu_{3{\rm V}}$ 
are correlated in pairwise 
---
if the values $F_{{\rm A}1}\equiv \mu_{1{\rm H}}/\mu_{1{\rm V}}$ and 
$F_{{\rm A}3}\equiv \mu_{3{\rm H}}/\mu_{3{\rm V}}$ are constant, they receive the two pairs in 
a pure state 
$\ket{\alpha_{12},\beta_{12}}\otimes \ket{\alpha_{34},\beta_{34}}$ 
with $\alpha_{12}/\beta_{12}=F_{{\rm A}1}$ and  
$\alpha_{34}/\beta_{34}=F_{{\rm A}3}$. 
If the values $\mu_{1{\rm H}}/\mu_{3{\rm H}}$ and $\mu_{1{\rm V}}/\mu_{3{\rm V}}$ are constant, 
Bob exchanges modes $1{\rm V}$ and $3{\rm H}$ before the 
transmission, and Alice exchanges modes back to obtain each pair 
in a pure state. The situation is similar for the case where the values 
$\mu_{1{\rm H}}/\mu_{3{\rm V}}$ and $\mu_{3{\rm H}}/\mu_{1{\rm V}}$ are 
constant. 

Let us consider an example in which Bob sends one member of a pair (mode $1$) 
to Alice through a polarization maintaining fiber and one member of the other 
pair (mode $3$) through the same fiber after a time delay $\Delta t$.  
Alice compensates the time delay $\Delta t$ after receiving the photons. 
The state $\ket{1}_{1{\rm H}}$, $\ket{1}_{3{\rm H}}$, $\ket{1}_{1{\rm V}}$, 
and $\ket{1}_{3{\rm V}}$ 
are transformed into $e^{i\phi_{{\rm H}}(t)}\ket{1}_{1{\rm H}}$, 
$e^{i\phi_{{\rm H}}(t+\Delta t)}\ket{1}_{3{\rm H}}$, 
$e^{i\phi_{{\rm V}}(t)}\ket{1}_{1{\rm V}}$, 
and $e^{i\phi_{{\rm V}}(t+\Delta t)}\ket{1}_{3{\rm V}}$, 
where $\phi_{{\rm H}}(t)$ and $\phi_{{\rm V}}(t)$ represent phase shifts 
in modes ${\rm H}$ and ${\rm V}$ induced by 
the fiber for photons input at time $t$. 
Since Bob initially has the photon pairs in the state 
$\ket{\Phi^{(+)}}_{12}\ket{\Phi^{(+)}}_{34}$, Alice and Bob share the 
photon pairs in the following states 
\begin{eqnarray}
& &\frac{e^{i[\varphi_+(t)+\varphi_-(t)]}}{\sqrt{2}}
	(\ket{1}_{1{\rm H}}\ket{1}_{2{\rm H}}
	+e^{-2i\varphi_-(t)}\ket{1}_{1{\rm V}}\ket{1}_{2{\rm V}}) \nonumber \\
	\otimes 
& &\frac{e^{i[\varphi_+(t+\Delta t)+\varphi_-(t+\Delta t)]}}{\sqrt{2}}
	(\ket{1}_{3{\rm H}}\ket{1}_{4{\rm H}} \nonumber \\
& &\qquad +e^{-2i\varphi_-(t+\Delta t)}\ket{1}_{3{\rm V}}\ket{1}_{4{\rm V}}), 
	\label{eq:last}
\end{eqnarray}
where $\varphi_+(t) \equiv [\phi_{{\rm H}}(t)+\phi_{{\rm V}}(t)] / 2$ and 
$\varphi_-(t) \equiv [\phi_{{\rm H}}(t)-\phi_{{\rm V}}(t)] / 2$. 
Assuming $\phi_{{\rm H}}(t)$ and $\phi_{{\rm V}}(t)$ are temporally 
fluctuating, this state becomes a mixed state. 
For simplicity, we assume the channel is symmetric about $H$ and $V$.  
The fluctuations of $\varphi_+(t)$ and $\varphi_-(t)$ are then independent, 
and we denote the correlation times of $\varphi_+(t)$ and $\varphi_-(t)$ 
by $\tau_+$ and $\tau_-$, respectively. 

We classify the situation into four cases: 
(a) $\Delta t \ll \tau_+, \tau_-$, 
(b) $\tau_+ \ll \Delta t \ll  \tau_-$, 
(c) $\tau_- \ll \Delta t \ll  \tau_+$, and 
(d) $\tau_+, \tau_- \ll \Delta t $. 
In case (a),  since we can use the approximations 
$\varphi_+(t)=\varphi_+(t+\Delta t)$ and $\varphi_-(t)=\varphi_-(t+\Delta t)$, 
if Bob exchanges modes $1{\rm V}$ and $3{\rm H}$ before the 
transmission and Alice exchanges back the modes, then they can share the 
maximally entangled photon pairs. 
Therefore it is not necessary to use the proposed purification scheme. 
In case (b), where 
$\varphi_+(t)\neq \varphi_+(t+\Delta t)$ and 
$\varphi_-(t)=\varphi_-(t+\Delta t)$, 
the above method does not work. But the proposed scheme works 
in this case as the mixture of Eq.~(\ref{eq:last}) has the form of 
Eq.~(\ref{eq:mixture}). 
In case (c), where $\varphi_+(t)=\varphi_+(t+\Delta t)$ and 
$\varphi_-(t)\neq \varphi_-(t+\Delta t)$, 
if Bob exchanges modes $1{\rm V}$ and  $3{\rm V}$ before the 
transmission and Alice exchanges back the modes, 
the situation is the same as in case (b). 
In case (d), where 
$\varphi_+(t)\neq \varphi_+(t+\Delta t)$ and 
$\varphi_-(t)\neq \varphi_-(t+\Delta t)$, 
Alice and Bob cannot obtain the photon pairs in the state of 
Eq.~(\ref{eq:mixture}), and they cannot purify the output 
even if the proposed scheme are used. 

In summary, we have proposed a purification scheme using 
linear optical elements and photon detectors. 
We have investigated errors in the output state when down-converted photons 
and imperfect detectors are used. 
We have shown that the errors can be discarded by the postselection because 
the error states contain less photons than 
the maximally entangled state. It became clear that the effect of 
dark counts is negligible. 
We have also discussed the required property of quantum channels for 
the proposed purification scheme. 

{\it Note added}. After the submission of this paper, 
another type of purification scheme for photon pairs was 
proposed by Pan {\it et al}.\cite{Zeilinger04}. 

\acknowledgements

We thank Kiyoshi Tamaki and Sahin K. \"Ozdemir for helpful discussions.  
This work was partly supported by a Grant-in-Aid for Encouragement of 
Young Scientists (Grant No. 12740243) and a Grant-in-Aid for Scientific 
Research (B) (Grant No. 12440111) by Japan Society of the Promotion of 
Science.

\end{multicols}

\end{document}